# Computable Phenotypes to Characterize Changing Patient Brain Dysfunction in the Intensive Care Unit

Yuanfang Ren, PhD[1,2], Tyler J. Loftus, MD[1,3], Ziyuan Guan, MS[1,2], Rayon Uddin, BS[1], Benjamin Shickel, PhD[1,2], Carolina B Maciel, MD[4], Katharina Busl, MD MS[4], Parisa Rashidi, PhD[1,5], Azra Bihorac, MD MS[1,2], Tezcan Ozrazgat-Baslanti, PhD[1,2]

[1] Intelligent Critical Care Center (IC3), University of Florida, Gainesville, FL, USA

[2] Department of Medicine, College of Medicine, University of Florida, Gainesville, FL, USA

[3] Department of Surgery, College of Medicine, University of Florida, Gainesville, FL, USA

[4] Department of Neurology, Neurocritical Care Division, College of Medicine, University of Florida, Gainesville, FL, USA

[5] Crayton Pruitt Family Department of Biomedical Engineering, University of Florida, Gainesville, FL, USA

Correspondence to: Azra Bihorac MD MS, Intelligent Critical Care Center (IC3)**,** Department of Medicine, Division of Nephrology, Hypertension, and Renal Transplantation, PO Box 100224, Gainesville, FL 32610-0224. Telephone: (352) 294-8580; Fax: (352) 392-5465; Email: abihorac@ufl.edu

Reprints will not be available from the author(s).

Funding/Support: Conflicts of Interest and Sources of Funding: A.B. and T.O.B. were supported by R01 grant GM110240 from the National Institute of General Medical Sciences. A.B. and T.O.B. were supported by Sepsis and Critical Illness Research Center P50 grant GM111152 from the National Institute of General Medical Sciences. A.B. was supported by a W. Martin Smith Interdisciplinary Patient Quality and Safety Award. T.O.B. received Clinical and Translational Science Institute grant 97071 and a grant from the Gatorade Trust, University of Florida. T.J.L. was supported by the National Institute of General Medical Sciences of the National Institutes of Health under K23 grant GM140268. R.P. was supported by the University of Florida Medical Student Summer Research fellowship. This work was supported in part by the National Institutes of Health and National Center for Advancing Translational Sciences Clinical and Translational Sciences Award UL1 grant TR000064. The content is solely the responsibility of the authors and does not necessarily represent the official views of the National Institutes of Health. The funders had no role in study design, data collection and analysis, decision to publish, or manuscript preparation. A.B. and T.O.B. had full access to all study data and take responsibility for data integrity and data analysis.

**Abstract**

**Background:** In the United States, more than 5 million patients are admitted annually to intensive care units, harboring an ICU mortality ranging from 10%-29% with costs exceeding $82 billion. Acute brain dysfunction status, delirium, often remains underdiagnosed or undervalued. The objective of this study was to develop automated computable phenotypes for acute brain dysfunction states and describe transitions among brain dysfunction states, which illustrate the clinical trajectories of ICU patients.

**Methods:** We created two single-center, longitudinal electronic health record datasets for 48,817 adult patients admitted to an ICU at UFH Gainesville (GNV) and Jacksonville (JAX). Each data source is inclusive of demographic information, comorbidities, vital signs, laboratory values, medications, and diagnoses and procedure codes for all index admissions as well as admissions within 12 months before and after index admissions. We developed algorithms to quantify acute brain dysfunction status including comatose (coma), delirium, normal, or death at 12-hour intervals of each ICU admission and to identify acute brain dysfunction phenotypes using continuous acute brain dysfunction status and *k*-means clustering approach.

**Results:** There were 49,770 admissions for 37,835 patients in the UFH GNV dataset and 18,472 admissions for 10,982 patients in the UFH JAX dataset. In total, 18% of patients had coma as the worst brain dysfunction status; every 12 hours, approximately 4%-7% would transit to delirium, 22%-25% would recover, 3%-4% would expire, and the remaining 67%-68% would remain in a coma in the ICU. Additionally, 7% of patients had delirium as the worst brain dysfunction status; approximately 6%-7% would transit to coma, 40%-42% would be no delirium, 1% would expire, and the remaining 51%-52% would remain delirium in the ICU. There were three phenotypes: persistent coma/delirium, persistently normal, and transition from coma/delirium to normal almost exclusively between first 48 hours after ICU admission.

**Conclusions**: We developed phenotyping scoring algorithms that determined acute brain dysfunction status every 12 hours while admitted to the ICU. This approach may be useful in developing prognostic and clinical decision-support tools to aid patients, caregivers, and providers in shared decision-making processes regarding resource use and escalation of care.

**INTRODUCTION**

Each year, 5.7 million patients are admitted to intensive care units (ICUs) in the United States, with ICU mortality ranging from 10%-29% and costs exceeding $82 billion, representing more than 4.1% of national health expenditures.[1, 2] A variety of scoring systems are utilized in ICUs; they are measures of disease severity that can be utilized to predict outcomes, provide an alert for impending worsening of the clinical status,[3] and support standardization of research and quality efforts. There are numerous prediction models for delirium, which is acute brain dysfunction with acute change or fluctuation from baseline mental status, inattention, and disturbance of awareness with varying degrees of cognitive impairment,[4] there are numerous prediction models. The models' performance varies,[5, 6] however, and delirium may remain frequently underdiagnosed or undervalued, in part likely because imputation of variables depends on nursing assessment and documentation.[7] Delirium and cognitive changes in the ICU course are highly relevant, as multiple associations with outcome including longer length of stay, increased mortality, and long-term cognitive outcome can ensue.[8]

On the other hand, most scoring systems require manual data collection and assessments that, although charted and processed via electronic medical records, contribute to ICU professionals' workload—which in turn may perpetuate alarm fatigue and burnout.[9, 10] This threat to the ICU workforce has only intensified with the effects of the pandemic.[11] As such, utilization of artificial intelligence to offload ICU workload is paramount.

The objective of this study was to develop and validate automated computable phenotypes for acute brain dysfunction states and describe transitions among brain dysfunction states, which illustrate the clinical trajectories of ICU patients. Our hypothesis is that automated phenotype computation is feasible and may allow clinicians to forego manually calculated scoring.

**METHODS**
**Data Source and Participant**

Using the University of Florida Health (UFH) Integrated Data Repository as Honest Broker, we have created two single-center, longitudinal datasets extracted directly from electronic health records (EHR) of all patients age 18 years or older admitted to UFH Gainesville (GNV) and Jacksonville (JAX). The UFH GNV dataset was searched to include 383,193 hospital admissions for 121,800 patients between June 1, 2014, and August 22, 2019. The UFH JAX dataset included 283,310 hospital admissions for 60,047 patients between June 1, 2014, and May 1, 2021. A patient was included in the UFH JAX cohort if any hospital admission of the patient is for surgery. The primary analyses focused on admissions with at least one ICU stay lasting at least 12 hours. ICU stays without any Richmond Agitation-Sedation Scale (RASS), Confusion Assessment Method (CAM), or Glasgow Coma Scale (GCS) measurement value were excluded (Figure 1). The UFH GNV dataset included 49,770 ICU admissions for 37,835 patients meeting this criterion, and the UFH JAX dataset included 18,472 ICU admissions for 10,982 patients meeting this criterion.

Each dataset included demographic information, comorbidities, vital signs, laboratory values, medications, and diagnoses and procedure codes for all index admissions as well as admissions within 12 months before and after index admissions. All EHR were de-identified, except that dates of service were maintained. This study was approved by the University of Florida Institutional Review Board and Privacy Office (IRB#201600223, IRB#201600262, and IRB#201901123) as an exempt study with a waiver of informed consent.

**Definition of Acute Brain Dysfunction States**

At the end of each 12-hour interval during each ICU stay, patient acute brain dysfunction status was classified as comatose (coma), delirium, normal, or death. Data elements used for identification of acute brain dysfunction states are provided in eTable 1. Computable phenotyping algorithms used a rule-based approach, as detailed below.

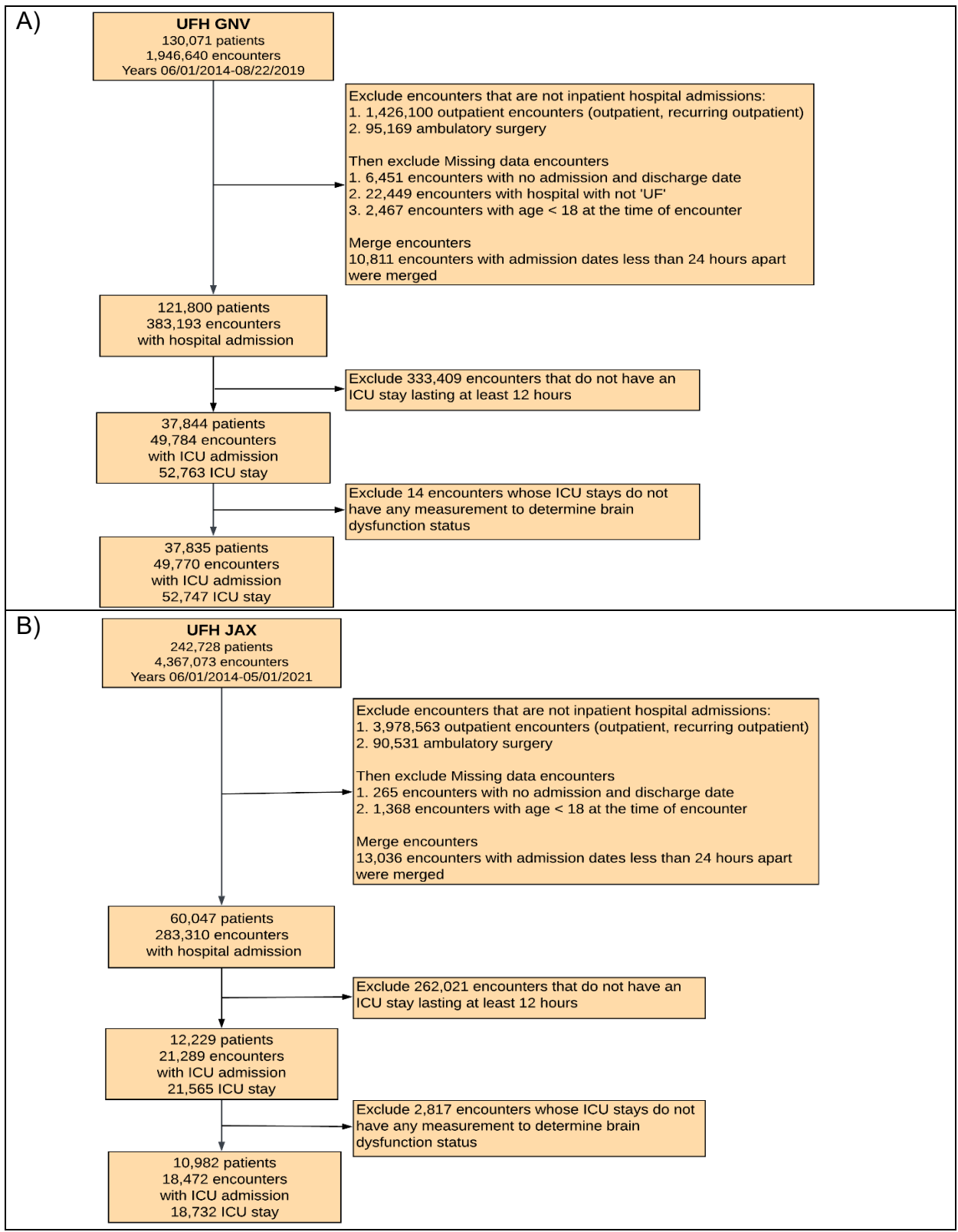

**Figure 1. Cohort selection and exclusion criteria.** (A) UFH GNV cohort (B) UFH JAX cohort.

**Identification of ICU Stay**

We identified the dates and times during which a patient was in an ICU bed using lists of ICU rooms and station data that show patient location at each time during hospitalization. Because patient movements within the hospital (e.g., to radiology and operating room suites) often generated separate electronic health record encounters though the patient was assigned to an ICU bed throughout those movements, ICU stays within 24 hours of one another were merged.

**Identification of Acute Brain Dysfunction States**

We identified acute brain dysfunction status using EHR data representing RASS, CAM, and GCS scores. At the end of each 12-hour interval, we determined the acute brain dysfunction states using the following rules ranked in the descending order of preference (Figure 2): (1) *coma* as defined if patients have at least one RASS score less than -3, or they have at least one RASS score equal to -3 and one temporally adjacent GCS score less than or equal to 8; (2) *delirium* as defined if patients have at least one RASS score equal to -3 and one temporally adjacent GCS score greater than 8, or they have at least one RASS score greater than -3 and one temporally adjacent CAM value with "positive" assessment; (3) *normal* as defined if patients have at least one RASS score greater than -3 and one temporally adjacent CAM value with "negative" assessment; (4) *missing* otherwise.

For intervals without sufficient scores to determine the acute brain dysfunction states (state is missing), missing values were imputed by forward propagating last available measurement in adjacent interval. CAM scores have three unique values including "positive," "negative," and "unable to assess," where the score with "unable to assess" was not used for imputation. If there were still intervals without sufficient scores to determine the acute brain dysfunction states, default values were imputed with RASS score equal to 0, CAM score equal to "negative," and GCS score equal to 15.

Patients dead in the ICU were determined as dead in the last interval. The date of death was determined using hospital records and the search of the Social Security Death Index.

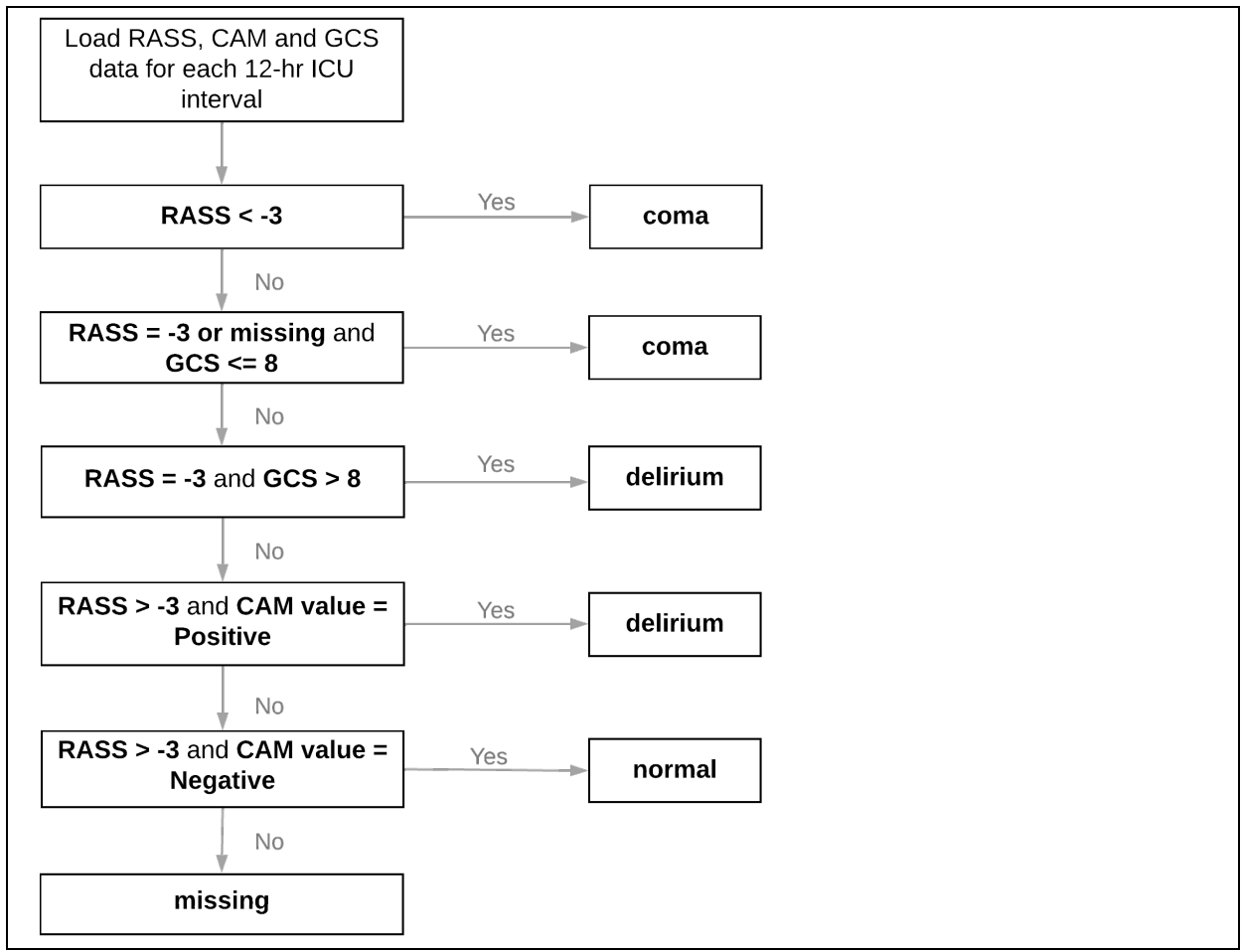

**Figure 2. Flowchart for phenotyping acute brain dysfunction states**

**Acute Brain Dysfunction States Patterns Derivation**

To understand how patient acuity brain dysfunction states evolve in the ICU, we used *k*-means clustering to derive acuity brain dysfunction states transition patterns within 3 days of ICU admission. For admissions lasting less than 3 days, we pad missing brain dysfunction status by carrying forward last available states. Thus, for each ICU admission, we had 6 (3 * 24 / 12) function states as clustering features.

**Statistical Methods**

We assessed frequency distributions and transition probability of acute brain dysfunction on UFH GNV and JAX cohorts. We conducted subgroup analysis using age (65 years or old), sex, race, and acuity score (a Sequential Organ Failure Assessment [SOFA] score above the median) to understand patient groups who would likely pose the greatest challenge in delirium assessment. We derived acute brain dysfunction states' patterns on UFH GNV dataset and validated the reproducibility of patterns by re-deriving clusters using the *k*-means clustering algorithm on UFH JAX dataset.

Data are represented as mean (SD) or median (IQR). For comparisons, we used the χ2 test for categorical variables and analysis of variance and the Kruskal-Wallis test for continuous variables. Analysis were performed with Python version 3.7.

## RESULTS
### Patients

The clinical characteristics and outcomes for patients were assessed for both cohorts (Table 1). UFH GNV and JAX datasets had similar distributions of age (59-60 years), sex (45%-47% female) and ethnicity, but significantly different race distribution (75% vs. 49% White, 19% vs. 46% African American). The median SOFA score for ICU patients was 3 for UFH GNV cohort and 2 for UFH JAX cohort; Modified Early Warning System (MEWS) score was 3 for UFH GNV cohort and 2 for UFH JAX cohort. UFH GNV cohort had longer median length of hospital stay than UFH JAX cohort (7 days vs. 4 days) but similar median length of ICU stay (4 days). Comparing to UFH JAX cohort, UFH GNV had the higher incidence of prolonged respiratory insufficiency (32% vs. 25% receiving mechanical ventilation, 54% vs. 25% of whom received more than 2 calendar days of ventilator support) and hospital mortality (9% vs. 3%).

**Table 1. Summary of patient characteristics.**

|  | UF Health GNV Admissions (06/01/2014-08/22/2019) | UF Health JAX Admissions (06/01/2014-05/01/2021) |
|---|---|---|
| Number of patients | 37,835 | 10,982 |
| Number of encounters | 49,770 | 18,472 |
| Number of ICU stays | 52,747 | 18,732 |
| **Preadmission clinical characteristics** | | |
| Age, mean (SD), years | 60 (17) | 60 (14) |
| Age greater than 65 years old, n (%) | 20,618 (41) | 6,714 (36) |
| Female sex, n (%) | 16,826 (44) | 5,108 (47) |
| Race, n (%)[a] | | |
|   White | 28,793 (76) | 5,574 (51) |
|   African American | 6,447 (17) | 4,877 (44) |
|   Other[b] | 1,974 (5) | 464 (4) |
|   Missing | 621 (2) | 67 (1) |
| Ethnicity, n (%)[a] | | |
|   Hispanic | 1,495 (4) | 320 (3) |
|   Non-Hispanic | 35,656 (94) | 10595 (96) |
|   Missing | 684 (2) | 66 (1) |
| **Acuity scores within 24h of ICU admission** | | |
| SOFA score, median (IQR) | 3 (2, 6) | 2 (1, 3) |
| MEWS score, median (IQR) | 3 (2, 4) | 2 (2, 3) |
| **Outcome** | | |
| Hospital days, median (IQR) | 7 (4, 13) | 4 (3, 9) |
| Days in ICU, median (IQR) | 4 (2, 7) | 4 (3, 6) |
| Mechanical ventilation, n (%) | 16,154 (32) | 4,601 (25) |
|   Mechanical ventilation calendar days, median (IQR) | 3 (2, 6) | 1 (1, 2) |
|   Mechanical ventilation greater than 2 calendar days, n (%) | 8,683 (54) | 1,138 (25) |
| Hospital mortality | 4,537 (9) | 486 (3) |

Abbreviation: SOFA: sequential organ failure assessment; MEWS: modified early warning score; ICU: intensive care unit; SD: standard deviation; IQR: interquartile range.

[a] Values were calculated per patient.

[b] Other race includes American Indian/Alaskan Native, Asian, Native Hawaiian/other Pacific Islander, and multiracial.

[c] Cardiovascular disease was considered if there was a history of congestive heart failure, coronary artery disease, or peripheral vascular disease.

**Distribution of Acuity Brain Dysfunction Status**

Distributions of brain dysfunction status for two cohorts within the initial 15 days of ICU admission are illustrated in Figure 3. In the UFH cohort, the number of ICU stays decreased rapidly from 50,000 to around 10,000 within 7 days, which marked the 75th percentile of ICU days. Brain dysfunction status was generally normal with few coma or delirium patients. The proportion of patients with coma was slightly higher than that with delirium. The proportion of patients with coma status decreased within first 48 hours after ICU admission while the proportion of patients with normal status increased. The maximum percentage of patients who were expired during each 12-hour period was 1%. Compared to UFH GNV cohort, UFH JAX cohort presented with more patients of normal status, and the average percentage of normal status was around 97% (UFH GNV cohort 83%). Similar to UFH GNV cohort, the proportion of patients with coma in UFH JAX cohort was slightly higher than that with delirium. The maximum percentage of patients who were expired during each 12-hour period was 0.5%.

Distributions of brain dysfunction status for two cohorts within the entire hospital stay are shown in Table 2. The worst brain dysfunction status over the entire hospital stay was selected as the final brain dysfunction status. Overall, 18% of patients had coma (UFH GNV 23% and UFH JAX 5%) and 7% of patients had delirium (UFH GNV 10% and UFH JAX 1%). Patients with coma or delirium occupied around median 17% of ICU stay (UFH GNV 17% and UFH JAX 13). Patients with coma had 28% hospital mortality, and patients with delirium had 9% hospital mortality.

**Table 2. Brain dysfunction status distribution.**

|  | Total (N = 68,242) | UF Health GNV Admissions (N = 49,770) | UF Health JAX Admissions (N = 18,472) | P-value |
|---|---|---|---|---|
| Coma, n (%) | 12,389 (18) | 11,508 (23) | 881 (5) | **0** |
| Percentage of intervals with coma status, %, median (IQR) | 17 (8, 38) | 17 (8, 39) | 14 (7, 30) | **<0.001** |
| Hospital mortality, n (%) | 3,525 (28) | 3,301 (29) | 224 (25) | **0.04** |
| Delirium, n (%) | 5,089 (7) | 4,921 (10) | 168 (1) | **0** |
| Percentage of intervals with delirium status, %, median (IQR) | 17 (8, 31) | 17 (9, 32) | 13 (6, 23) | **<0.001** |
| Hospital mortality, n (%) | 450 (9) | 436 (9) | 14 (8) | 0.92 |
| Coma or delirium, n (%) | 17,478 (26) | 16,429 (33) | 1,049 (6) | **0** |
| Percentage of intervals with coma or delirium status, %, median (IQR) | 17 (8, 31) | 17 (9, 32) | 13 (6, 23) | **<0.001** |
| Hospital mortality, n (%) | 3,975 (23) | 3,737 (23) | 238 (23) | 1 |
| Normal, n (%) | 50,764 (74) | 33,341 (67) | 17,423 (94) | 0 |
| Hospital mortality, n (%) | 1,048 (2) | 800 (2) | 248 (1) | **<0.001** |

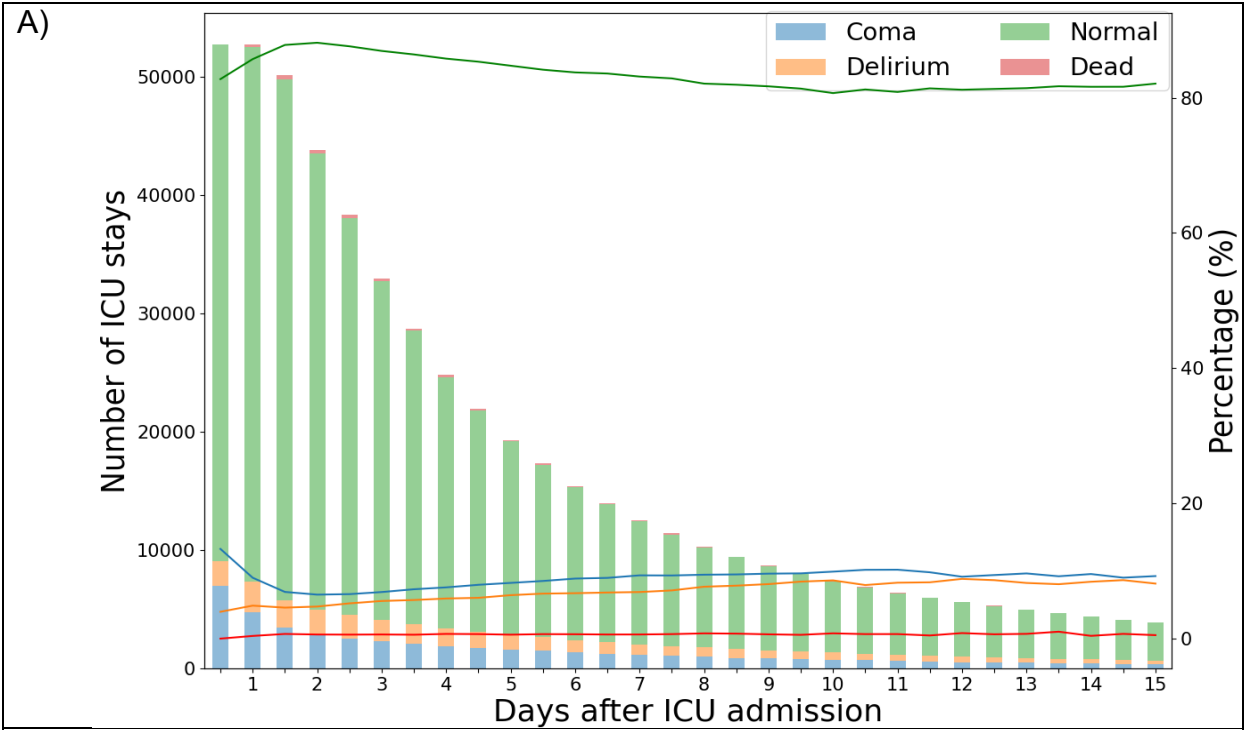
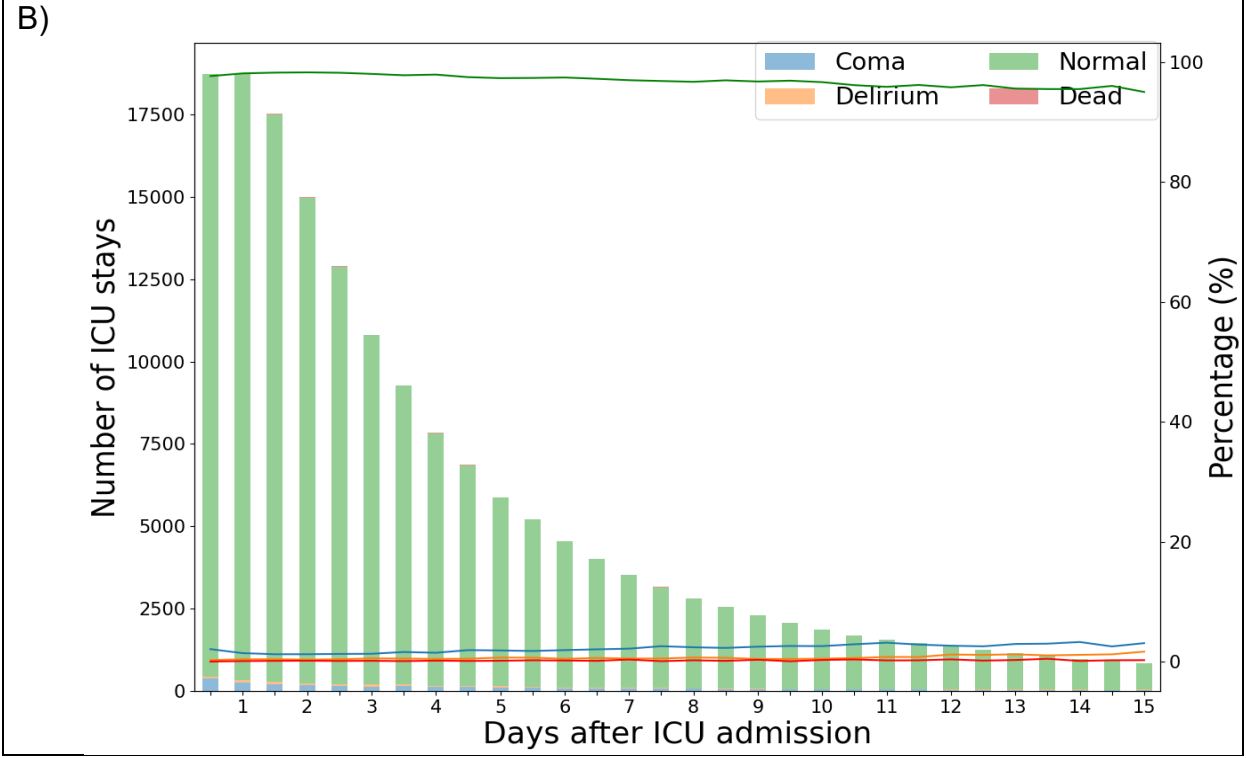

**Figure 3. Distribution of brain dysfunction status within 15 days of ICU admission.** (A) UFH GNV cohort (B) UFH JAX cohort. The left y-axis shows the total number of ICU stays for all brain dysfunction states. The right y-axis shows the percentage of each brain dysfunction state.

**Subgroup Analysis**

Distributions of acuity brain dysfunction for patient subgroups were illustrated in Table 3. Subgroup analysis was conducted among patient subgroups based on age, sex, race, and SOFA score. There was a significantly higher proportion of patients with coma or delirium for older patients with age greater than 65 years older (27% vs. 24% for younger patients, p<0.001); male patients (23% vs. 21% for female patients, p<0.001); white patients (28% vs. 20% for African American patients vs. 26% for other patients, p<0.001); and patients with higher SOFA score greater than 3 (45% vs. 11% for patients with lower SOFA score). Patients with higher SOFA score also exhibited longer median time coma or delirium status compared to patients with lower score (25% vs. 15%). Out of patients with coma or delirium, those with either older age or higher SOFA score had significantly higher hospital mortality compared to those with younger age (28% vs. 19%, p<0.001) or lower SOFA score (26% vs. 14%, p<0.001).

**Transition Probability Matrix of Acuity Brain Dysfunction Status**

Acuity brain dysfunction status transition probabilities are illustrated in Figure 4. For patients with coma, every 12 hours, approximately 4%-7% would transit to delirium, 22%-25% would recover, 3%-4% would expire, and the remaining 67%-68% would remain comatose in the ICU. For patients with delirium, every 12 hours, approximately 6%-7% would transit to coma, 40%-42% would be no delirium, 1% would expire, and the remaining 51%-52% would remain delirium in the ICU.

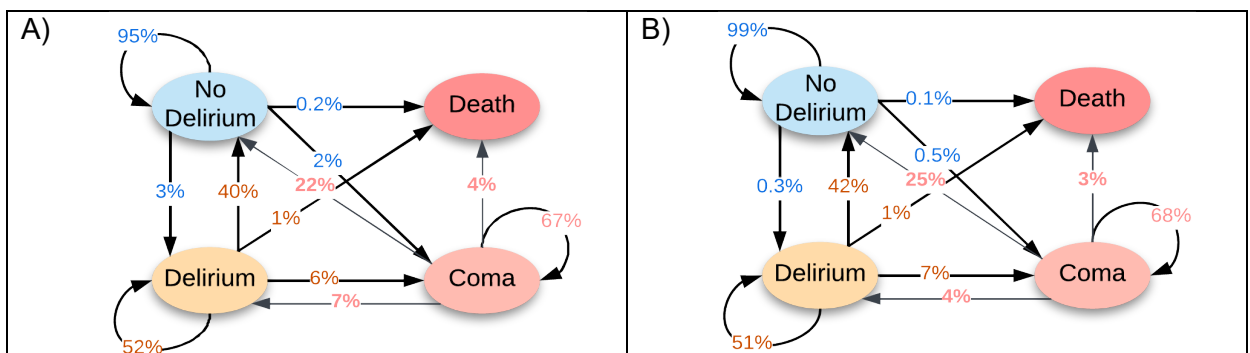

**Figure 4. Transitions between acute brain dysfunction states in every 12-hour interval in ICU.** (A) UFH GNV cohort (B) UFH JAX cohort.

**Table 3. Brain dysfunction status distribution of patient subgroups.** Patients are grouped based on age, sex, race, and SOFA score.

| | Age | | | Sex | | | Race | | | | SOFA | | |
|---|---|---|---|---|---|---|---|---|---|---|---|---|---|
| | Age <= 65 (N = 40,910, 60%) | Age > 65 (N = 27,332, 40%) | P-value | Female (N = 30,976, 45%) | Male (N = 37,266, 55%) | P-value | White (N = 45,917, 67%) | African American (N = 18,431, 27%) | Other (N = 3,165, 5%) | P-value | SOFA <= 3 (N = 39,111, 57%) | SOFA > 3 (N = 29,131, 43%) | P-value |
| Coma, n (%) | 7,237 (18) | 5,152 (19) | <0.001 | 5,124 (16) | 7,265 (19) | <0.001 | 8,747 (19)[a] | 2,813 (15) | 594 (19)[a] | <0.001 | 2,127 (5) | 10,262 (35) | 0 |
| Percentage of intervals with coma status, %, median (IQR) | 17 (8, 38) | 17 (8, 38) | 0.74 | 17 (8, 38) | 17 (8, 38) | 0.97 | 17 (8, 34)[a] | 17 (8, 40) | 15 (8, 40) | 0.02 | 10 (5, 20) | 18 (9, 43) | <0.001 |
| Hospital mortality, n (%) | 1,717 (24) | 1,808 (35) | <0.001 | 1,545 (30) | 1,980 (27) | <0.001 | 2,518 (29)[a] | 743 (26) | 144 (24) | 0.006 | 445 (21) | 3,080 (30) | <0.001 |
| Delirium, n (%) | 2,725 (7) | 2,364 (9) | <0.001 | 2,316 (7) | 2,773 (7) | 0.87 | 3,881 (8)[a] | 923 (5) | 219 (7)[a] | <0.001 | 2,143 (5) | 2,946 (10) | <0.001 |
| Percentage of intervals with delirium status, %, median (IQR) | 15 (8, 29) | 17 (9, 33) | <0.001 | 16 (8, 31) | 17 (9, 31) | 0.32 | 17 (9, 33) | 16 (8, 29) | 14 (8, 25) | 0.03 | 14 (7, 28) | 17 (9, 33) | <0.001 |
| Hospital mortality, n (%) | 168 (6) | 282 (12) | <0.001 | 211 (9) | 239 (9) | 0.57 | 343 (9) | 80 (9) | 16 (7) | 0.73 | 142 (7) | 308 (10) | <0.001 |
| Coma or delirium, n (%) | 9,962 (24) | 7,516 (27) | <0.001 | 7,440 (24) | 10,038 (27) | <0.001 | 12,628 (28)[a] | 3,736 (20) | 813 (26)[a] | <0.001 | 4,270 (11) | 13,208 (45) | 0 |
| Percentage of intervals with coma or delirium status, %, median (IQR) | 22 (11, 44) | 23 (11, 46) | 0.02 | 21 (11, 43) | 23 (11, 47) | 0.07 | 22 (11, 44) | 22 (11, 48) | 20 (10, 43) | 0.09 | 15 (8, 30) | 25 (13, 50) | <0.001 |
| Hospital mortality, n (%) | 1,885 (19) | 2,090 (28) | <0.001 | 1,756 (24) | 2,219 (22) | 0.02 | 2,861 (23) | 823 (22) | 160 (20) | 0.12 | 587 (14) | 3,388 (26) | <0.001 |

Abbreviation: SOFA: sequential organ failure assessment; IQR: interquartile range.

**Phenotypes of Acuity Brain Dysfunction Status**

Clustering analyses of the UFH GNV cohort (Figure 5a) and the UFH JAX cohort (Figure 5b) suggested three acuity brain dysfunction status phenotypes during the first 72 hours of ICU admission. Cluster 1 manifested as persistent normal status with the largest proportion of patients (UFH GNV 40,814, 82% and UFH JAX 17,973, 97%), lowest in-hospital mortality (UFH GNV 5% and UFH GNV 2%), and shortest median lengths of stay in the ICU (UFH GNV 3 days and UFH JAX 4 days) and hospital (UFH GNV 7 days and UFH JAX 4 days). Cluster 2 initially presented with coma/delirium brain status but then transitioned to normal, and this occurred almost exclusively within the 48 hours of ICU admission. This cluster had the intermediate proportion of patients (UFH GNV 5,329, 11% and UFH JAX 268, 1%), intermediate in-hospital mortality (UFH GNV 10% and UFH GNV 7%), and longer median lengths of stay in the ICU (UFH GNV 6 days and UFH JAX 6 days) and hospital (UFH GNV 10 days and UFH JAX 8 days). Cluster 3 was characterized by persistent coma/delirium status with the smallest proportion of patients (UFH GNV 3,627, 7% and UFH JAX 231, 1%), the highest in-hospital mortality (UFH GNV 52% and UFH GNV 39%), and longer median lengths of stay in the ICU (UFH GNV 6 days and UFH JAX 5 days) and hospital (UFH GNV 9 days and UFH JAX 10 days).

**DISCUSSION**

Using EHR data, we have developed a computable phenotyping approach to identify and characterize consistent acute brain dysfunction status. We found that within first 48 hours after ICU admission, there was an increase in the raw number and percentage of ICU patients without brain dysfunction status. For patients with coma/delirium brain dysfunction status, there was a 32%/48% probability that the acute brain dysfunction status would be different 12 hours later. The probability of death was three-fold higher among coma vs. delirium patients and eleven-fold higher among coma/delirium vs. normal patients, suggesting that our methods for classifying brain dysfunction status were effective in identifying patients at risk for death. Patients with older age or higher SOFA score were associated with increased risk for coma or delirium status as well as death, consistent with risk factor selection for delirium prediction model.[12] Finally, through clustering analysis, we identified three ICU patient phenotypes: patients who were persistently normal, patients who persistently had coma/delirium status, and patients who initially had brain dysfunction but became normal within 48 hours.

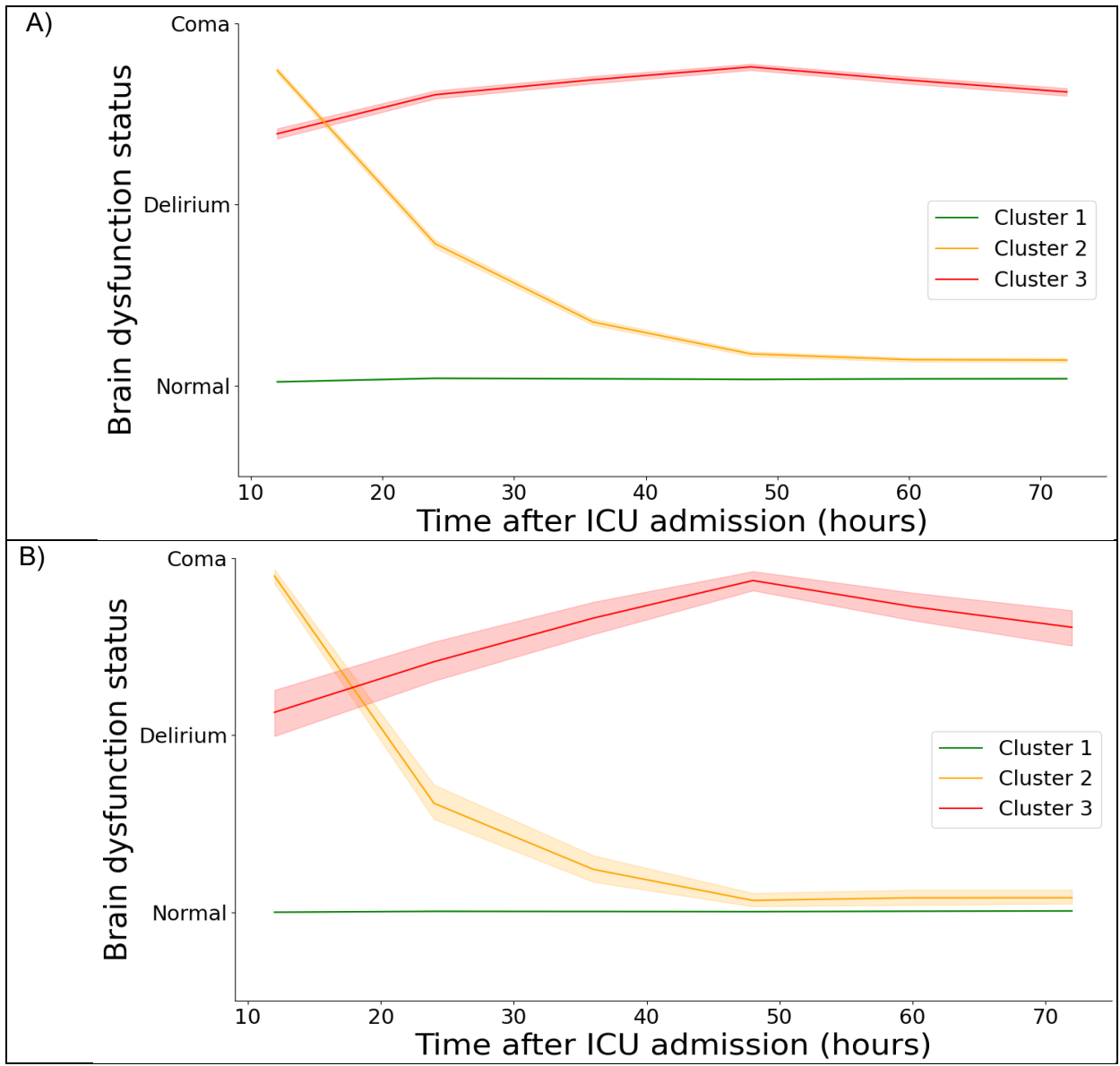

**Figure 5. Acuity brain dysfunction status distribution across phenotypes within first three days of ICU admission.** (A) UFH GNV cohort (B) UFH JAX cohort.

Such observations involving ICU patient phenotypes may inform prognostic discussions among patients, caregivers, and providers. At the time of ICU admission for patients with brain dysfunction status, patients and their caregivers may wish to embark on a course of aggressive, life-sustaining treatments if there is a high probability of recovery and transition to normal status. Some critically ill patients express or have previously expressed a desire to forego prolonged life-sustaining treatments. In these cases, there may be utility in providing the patient and their caregivers with accurate, data-driven predictions that the probability of early recovery is low. This information could augment the decision-making process and alleviate the stress associated with the decision to forgo aggressive, resource-intense therapy.

Computable phenotypes using established data standards and common data models provide the opportunity to get the fast and accurate annotation of acute brain dysfunction status across multiple institutions. Automated, consistent, and accurate identification of acute brain dysfunction status using EHR data further lays the foundation of developing artificial intelligence applications to facilitate early recognition and appropriate management with targeted preventative and therapeutic interventions for ICU delirium patients.[13] The authors are unaware of any computable phenotype that identifies and characterizes acute brain dysfunction status using EHR data that can be easily adapted to different data models and further used in real time.

This study has limitations. First, our use of data from UFH JAX did not include all hospital admissions, which limits the generalizability. Second, our result lacks validation. Finally, whether such identification of acute brain dysfunction improves prognostication and clinical decision-making remains unknown.

**CONCLUSION**

We developed phenotyping algorithms that determined patient acuity brain dysfunction status every 12 hours during ICU admission. This task can be automated, which has the advantage of avoiding additional patient assessments by ICU health care workers, who already face worsening workforce shortages and job-related stress and burnout. Automated phenotyping has the potential to leverage high-resolution physiological signals and digital EHR data to develop prognostic and clinical decision-support tools that aid patients, caregivers, and providers in shared decision-making processes regarding resource use and escalation of care that is consistent with patient values.

**eTable 1. Data elements in the acuity brain dysfunction states algorithm**

| Used for identification | Features | Description | Format |
|---|---|---|---|
| Identifier | patient_deiden_id | Deidentified Patient ID | Strings |
| Identifier | encounter_deiden_id | Deidentified Patient ID | Strings |
| RASS | meas_value | Measured Value | Measured Value |
| RASS | recorded_time | Recorded Time | The date and time value was recorded |
| RASS | vital_sign_measure_name | Name of the measured variable | Name of the measured variable |
| CAM | meas_value | Measured Value | Measured Value |
| CAM | recorded_time | Recorded Time | The date and time value was recorded |
| CAM | disp_name | Displayed name | Strings |
| CAM | vital_sign_measure_name | Name of the measured variable | Name of the measured variable |
| GCS | glasgow_coma_adult_score | Measured Value | Measured Value |
| GCS | glasgow_coma_datetime | Recorded Time | The date and time value was recorded |